\documentclass[runningheads,a4paper]{llncs}

\usepackage{amssymb}
\usepackage{graphicx}

\usepackage{url}

\urldef{\mailsa}\path|{sudharakap, usefi}@mun.ca|  
  
\newcommand{\keywords}[1]{\par\addvspace\baselineskip
	\noindent\keywordname\enspace\ignorespaces#1}

\usepackage{times}

\usepackage{caption}
\usepackage{cite}

\usepackage[cmex10]{amsmath}

\usepackage{marvosym}
\usepackage{csquotes}
\usepackage{amssymb}
\usepackage{multirow}
\usepackage{mathtools}

\usepackage{varioref}

\usepackage{caption}
\begin{document}
\pagenumbering{gobble}

\title{Homomorphic Evaluation of Database Queries}

\titlerunning{Homomorphic Evaluation of Database Queries}

\author{Sudharaka Palamakumbura and Hamid Usefi\thanks{The research is supported by NSERC of Canada under grant \# RGPIN 418201 and the Research \& Development Corporation of Newfoundland and Labrador.}}
\institute{Department of Mathematics and Statistics\\
Memorial University of Newfoundland\\
St. John's, NL, Canada, A1C 5S7\\
\mailsa }

\maketitle

\begin{abstract}

Homomorphic encryption is an encryption method that enables computing over encrypted data. This has a wide range of real world ramifications such as being able to blindly compute a search result sent to a remote server without revealing its content. This paper discusses how database search queries can be made secure using a homomorphic encryption scheme.  We propose a new database search technique that can be used with the ring-based fully homomorphic encryption scheme proposed by Braserski. 
\keywords{homomorphic; privacy; encryption; database; query}
\end{abstract}

\section{Introduction}

According to a  recent study, 74\% of smartphone users use a location based service (such as Google Maps) to find directions and other location based information \cite{Kathryn}. Moreover, the adaptation of these kind of services in healthcare are becoming increasingly common with cloud-based health recording and genomic data management tools such as Microsoft Health. However, the widespread adaptation of location-based services poses a threat to users because their personal data, such as location, health records, and sometimes even genomic data, is shared on the web without any guarantee of privacy. 

In this work, we address the problem of searching privately on a database. We consider the scenario that a user wants to  send a search request to a server and would want the server to learn nothing about his query. So, it makes sense that the user encrypts his search using his public key and sends the cipher-text over to the server. We consider the case where data over the server is not encrypted.
This applies in particular to queries sent to search engines. The case where the data over the server is encrypted will be treated differently elsewhere. Our proposed scheme shall use a homomorphic encryption scheme. 

Homomorphic encryption  allows computations to be carried out on the cipher-text such that after decryption, the result would be the same as carrying out identical computations on the plain-text. This has novel implications such as being able to carry out operations on database queries in the form of cipher-text and returning the result to the user so that no information about the query is revealed at the server's end\cite{Boneh}. 

The idea of homomorphic encryptions is not new, and even the oldest of ciphers, ROT13 developed in ancient Rome, had homomorphic properties with respect to string concatenations\cite{Hesse}. Certain modern ciphers such as RSA  and El Gamal also support homomorphic multiplication of cipher texts\cite{Hesse}. 

The idea of a ``fully" homomorphic encryption scheme (or \textit{privacy homomorphism}) which supports two homomorphic operations was first introduced by Rivest, Adleman, and Dertouzous in 1978\cite{Rivest}. After more than three decades, the first fully homomorphic encryption scheme was founded by Gentry in 2009 with his breakthrough construction of a lattice based cryptosystem that supports both homomorphic additions and multiplications\cite{homenc}. Although the lattice based system is not used in practice, it paved the way for many other simpler and more efficient fully homomorphic models constructed afterwards. 

At a high level, Gentry's idea can be described by the following general model. This is the blueprint that is used in all homomorphic encryption schemes that followed. 

\begin{enumerate}
	\item Develop a \textit{Somewhat Homomorphic Encryption Scheme} that is restricted to evaluating a finite number of additions or multiplications.
	\item Modify the somewhat homomorphic encryption scheme to make it \textit{Bootstrappable}, that is, modifying it so that it could evaluate its own decryption circuit plus at least one additional NAND gate. 
\end{enumerate}
Every probabilistic encryption function usually introduces a \textit{noise}   and when the noise exceeds a certain threshold, the decryption function does  not return the desired plain-text. The idea behind constructing a bootstrappable scheme is that whenever the noise level is about to reach the threshold, we can \textit{bootstrap} the cipher-text and get a new cipher-text so that   these cipher-texts decrypt to the same pain-text but the new  cipher-text will have  a lower noise. In this way, if the cipher-text is bootstrapped from time to time, an arbitrary number of operations can be carried out.

Our work improves upon a method proposed by Gahi et al.\cite{Gahi[2]} to homomorphically encrypt database queries. Their work specifically uses the DGHV fully homomorphic encryption scheme \cite{Dijk}. The DGHV scheme operates on plain-text bits separately, and thus Gahi's method requires a large amount of computations to perform even on a simple operation such as integer multiplication. We propose an alternative to Gahi's method, which we call \textit{Homomorphic Query Processing}. Our method is not restricted to the DGHV scheme and can be used with more modern fully homomorphic encryption schemes. For example, using our Homomorphic Query Processing technique with the more recent ring based fully homomorphic encryption scheme proposed by Braserski et al.\cite{Braserski}, which work on blocks of data (such as integers) rather than single bits (as in Gahi's scheme), the number of computations can be greatly reduced.

\section{DGHV Fully Homomorphic Encryption}
The DGHV scheme was introduced by Marten van Dijk, Craig Gentry, Shai Halevi, and Vinod Vaikuntanathan in 2010, and this scheme operates on integers as opposed to lattices in Gentry's original construction. The scheme follows Gentry's original blueprint by first constructing a somewhat homomorphic encryption scheme. The key generation, encryption and decryption algorithms of the DGHV scheme are given below. 

Let $\lambda\in\mathbb{N}$ be the security parameter and set $N=\lambda,\,P=\lambda^2$ and $Q=\lambda^5$. The scheme is based on the following algorithms;
\begin{itemize}
\item \textbf{KeyGen($\lambda$)}: The key generation algorithm that randomly chooses a $P$-bit integer $p$ as the secret key. 
\item \textbf{Enc($m,p$)}: The bit $m\in\{0,1\}$ is encrypted by 
\[c=m'+pq,\] 
where $m'\equiv m\mbox{ (mod 2)}$ and $q,\,m'$ are random $Q$-bit and $N$-bit numbers, respectively. Note that we can also write the cipher-text as $c=m+2r+pq$ since $m'=m+2r$ for some $r\in\mathbb{Z}$.
\item \textbf{Dec($c,p$)}: Output $(c\mbox{ mod p})\mbox{ mod 2}$ where $(c\mbox{ mod p})$ is the integer $c'$ in $(-p/2,p/2)$ such that $p$ divides $c-c'$.
\end{itemize}
The value $m'$ is called the \textit{noise} of the cipher-text. Note that this scheme, as it is given above, is symmetric (i.e., it only has a private key). We can define the public key as a random subset sum of encryptions of zeros, that is, the public key is a randomly choosen sum from a predefined set of encryptions of zeros: $S=\{2r_1+pq_1,2r_2+pq_2,\ldots,2r_n+pq_n\}$. A typical encryption of the plain-text $m$ would be,
\begin{eqnarray*}c&=&m+\sum_{i\in T}(2r_i+pq_i)\\
&=&m+2\sum_{i\in T}r_i+p\sum_{i\in T}q_i,
\end{eqnarray*}
where $T\subseteq S$. From here on we shall use $m'$ to denote $m+\sum_{i\in T}r_i$ and $q$ to denote $\sum_{i\in T}q_i$.

This scheme is homomorphic with respect to addition and multiplication and decrypts correctly as long as the noise level does not exceed $p/2$ in absolute value. That is, $\left|m'\right|<p/2$. Hence, this is a somewhat homomorphic encryption scheme in the sense that once the noise level exceeds $p/2$, the scheme loses its homomorphic ability. It is shown that this scheme is Bootstrappable.

\section{Query Processing Using the DGHV Scheme}
The DGHV scheme can be used to create a protocol that establishes blind searching in databases. This method was proposed by Gahi et al.\cite{Gahi[2]}.

Suppose we need to retrieve a particular record from the database. Typically, we send a query to the database encrypted using the DGHV scheme. Let $v_i$ be the $i$\textsuperscript{th} bit of the query $v$ and $c_i$ be the $i$\textsuperscript{th} bit of a record $R$ in database $D$. Both the query and the database record is encrypted using the DGHV scheme. Suppose the plain-text bit corresponding to $v_i$ is $m_i$ and the plain-text bit corresponding to $c_i$ is $m'_i$. Then,
\[v_i=m_i+2r_i+pq_i\]
and
\[c_i=m'_i+2r'_i+pq'_i,\]
where $r_i,r'_i,q_i\mbox{ and }q'_i$ are random numbers and $p$ is the secret key. The server shall compute the following sum for each record $R_t$  with index $t$:
\begin{equation}
I_t=\prod_{i}(1+c_{i}+v_{i}).\label{GahiEquation}\end{equation}

\begin{figure}[h]
\center
\includegraphics[width=1\textwidth]{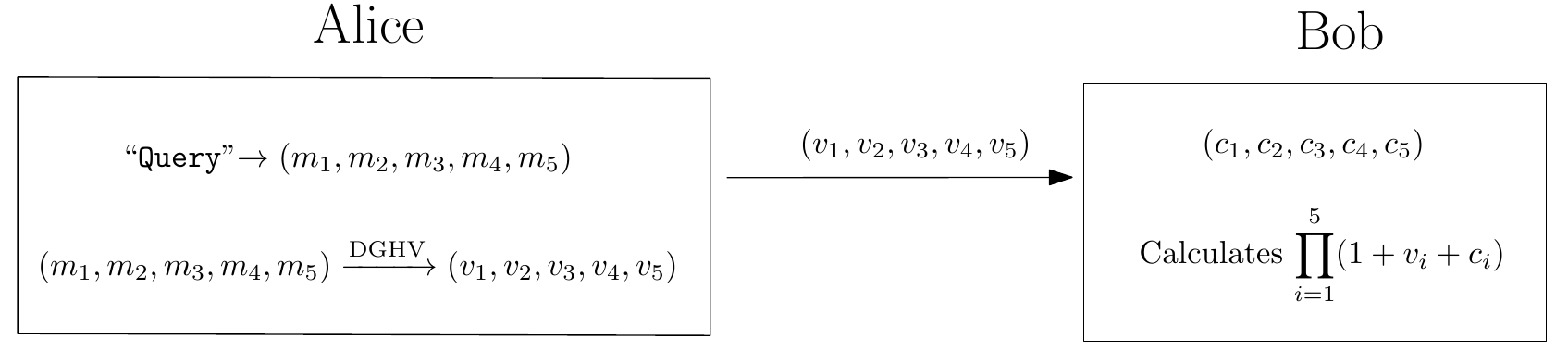}
\caption{Calculation of $I_r$ values.\label{figure6}}
\end{figure}

We observe that 
\begin{eqnarray*}
1+c_i+v_i
&=&1+(m_i+m'_i)+2(r_i+r'_i)+p(q_i+q'_i).
\end{eqnarray*}
So, if $m_i=m'_i$, then $m_i+m'_i\equiv 0 \mod 2 $. In this case:
\begin{eqnarray*}
1+c_i+v_i
&=&\mbox{Enc}(1).
\end{eqnarray*}
On the other hand, if $m_i\neq m'_i$, then $m_i+m'_i\equiv 1 \mod 2$. Therefore,
\begin{eqnarray*}
1+c_i+v_i
&=&2(1+r_i+r'_i)+p(q_i+q'_i)\\
&=&\mbox{Enc}(0).
\end{eqnarray*}
This results in $I_t=\mbox{Enc}(0)$. Hence, for each record $R_t$ in the database we will have an $I_t$ value that is equal to $\mbox{Enc}(1)$ or $\mbox{Enc}(0)$ depending on whether the search query $m$  matches $R_t$ or not.

 Next, we calculate the partial sums of the $I_t$ values: 
\begin{align}
S_r=\sum_{t\leq r}I_{t}. \label{PartialSums}
\end{align}

As an example, let us consider a database that has five records, each encoded with $4$ bits. If the query sent by the user is $(\mbox{Enc}(1),\mbox{Enc}(1),\mbox{Enc}(0),\mbox{Enc}(0))$, we obtain the corresponding $I_r$ and $S_r$ values, as shown in Table \ref{table1}. 

\begin{table}[ht]
\centering
\begin{tabular}{|c||c||c|}
\hline
Database Records & $I_r$& $S_r$\\
\hline
$(1,1,0,0)$ & $\mbox{Enc} (1)$&\mbox{Enc} (1)\\
$(1,0,1,0)$ & $\mbox{Enc} (0)$&\mbox{Enc} (1)\\
$(1,1,0,0)$ & $\mbox{Enc} (1)$&\mbox{Enc} (2)\\
$(1,1,0,1)$ & $\mbox{Enc} (0)$&\mbox{Enc} (2)\\
$(1,0,0,0)$ & $\mbox{Enc} (0)$&\mbox{Enc} (2)\\
\hline
\end{tabular}
\caption{Sample database  with corresponding $I_r$ and $S_r$ values \label{table1}}
\end{table}

Next, we   calculate the sequence $(I'_{r,j})$ for every record $R_r$ with index $r$ and every positive integer $j\leq r$:

\begin{equation}I'_{r,j}=I_r\prod_{i}(1+\bar j_i+S_{r,i})\label{GahiEquation2},\end{equation}
where $S_{r,i}$ is the $i$\textsuperscript{th} bit of $S_r$ and $\bar j_i$ represents the $i$\textsuperscript{th} bit of the  encryption of $j$. Hence, these sequences have the property that whenever $I_r=\mbox{Enc}(1)$ and $S_r=\mbox{Enc}(j)$, we have $I'_{r,j}=\mbox{Enc}(1)$. Otherwise, $I'_{r,j}=\mbox{Enc}(0)$. Following the  example given in Table \ref{table1}, we get

\begin{align*}
(I'_{1})&=(\mbox{Enc}(1)),\\
(I'_{2})&=(\mbox{Enc}(0),\mbox{Enc}(0)),\\
(I'_{3})&=(\mbox{Enc}(0),\mbox{Enc}(1),\mbox{Enc}(0)),\\
(I'_{4})&=(\mbox{Enc}(0),\mbox{Enc}(0),\mbox{Enc}(0),\mbox{Enc}(0)),\\
(I'_{5})&=(\mbox{Enc}(0),\mbox{Enc}(0),\mbox{Enc}(0),\mbox{Enc}(0),\mbox{Enc}(0)).
\end{align*}

\begin{figure}[h]
\center
\includegraphics[width=1\textwidth]{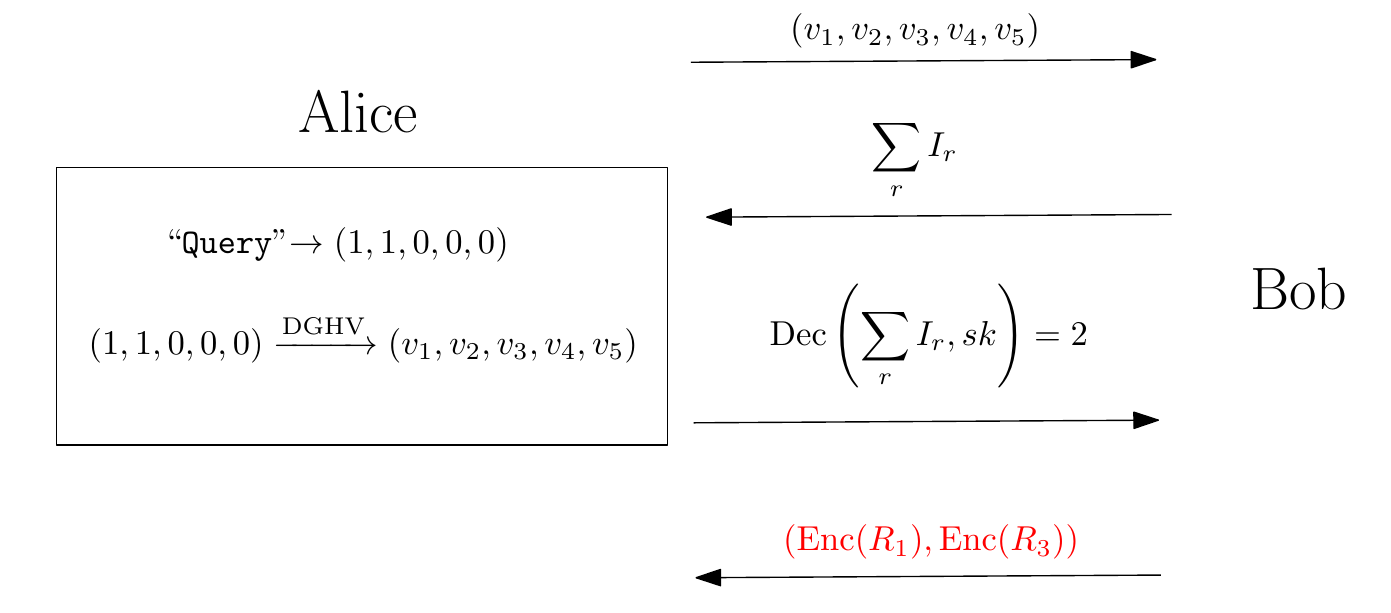}
\caption{Alice, Bob, and Gahi's Protocol.\label{Alice[2]}}
\end{figure}

Finally, we calculate, 
\begin{equation}(R')=\sum_{k}\mbox{Enc}(R_{k})(I'_{k}),\label{GahiEquation3}\end{equation}
where $R_{k}$ is the $k$\textsuperscript{th} record in $D$. So, $(R')$ is a sequence containing only the encrypted records that matches our search query. Note that the definition of $(R')$ relies on adding vectors of different lengths. This is done in the natural way, whereby all the vectors are made the same length by padding with zeros prior to addition. In the above example, we obtain,
\[(R')=(\mbox{Enc}(R_{1}),\mbox{Enc}(R_{3}),\mbox{Enc}(0),\mbox{Enc}(0)).\]
At this point, the sequence $(R')$ will contain all the records that match our query, but with trailing encryptions of zeros we do not need. Hence, a second sum is calculated at the server side to determine the number of terms that are useful in the sequence: 
\[n=\sum_{r} I_r\]
This result can be returned to the user and decrypted to obtain the number of records that match the search query. Hence, the sequence $(R')$ can be truncated at the appropriate point and returned to the user for decryption. The whole process is illustrated in Figure \ref{Alice[2]}.

An update query can be performed by,
\[R_{new}=(1+I_{r})R+I_{r}U, \mbox{ for every } R\in D,\]
where $U$ is the new value that we wish to insert whenever the query matches $R\,(\mbox{or }I_r=\mbox{Enc}(1))$. 
A deletion of a record can be performed by,
\[R_{new}=(1+I_{r})R\mbox{ for every } R\in D.\]
To perform all these operations without exceeding the maximum noise permitted ($p/2$), it is necessary to choose the parameters $N, P,\mbox{ and }Q$ appropriately. 

Gahi's method works on plain-text bits and thus requires significant computational ability on the part of the server. This is due to the fact that it is restricted to the DGHV scheme which processes plain-text bits separately. Now we propose an alternative protocol called the Homomorphic Query Processing Scheme. This protocol enables us to process database queries using more modern fully homomorphic encryption schemes such as the ring based scheme proposed by Braserski et al.\cite{Braserski}, which acts on blocks of plain-text rather than single bits. 
\section{Homomorphic Query Processing}
The main drawback in Gahi's method is that it requires an enormous number of homomorphic operations because it employs the DGHV encryption scheme, which uses bitwise encryption. We propose an alternative protocol called \textit{Homomorphic Query Processing} that is compatible with the more recent ring-based fully homomorphic encryption scheme introduced by Braserski et al.\cite{Braserski}. The major advantage is that Braserski's method works on plain-text and cipher-text blocks and thus the number of homomorphic operations required can be greatly reduced.

We first give a brief introduction to the ring based fully homomorphic Encryption Scheme proposed by Braserski, and then proceed to define our Homomorphic Query Processing method.

\subsection{Ring Based Fully Homomorphic Encryption}
This encryption scheme was introduced by Braserski, et al \cite{Braserski} and operates on the polynomial ring $R=\mathbb{Z}[X]/\left<f(x)\right>$; the ring of polynomials with integer coefficients modulo $f(x)$, where,
\[f(x)=\prod_{\substack{1\leq k\leq n\\\gcd(k,n)=1}}\left(x-e^{2i\pi\frac{k}{n}}\right),\] is the $n$\textsuperscript{th} cyclomatic polynomial. The plain-text space is the ring $R_t=\mathbb{Z}_t[x]/\left<f(x)\right>$, where $t$ is an integer. The key generation and encryption functions make use of two distributions $\chi_{key}$ and $\chi_{err}$ on $R$ for generating small elements. The uniform distribution $\chi_{key}$ is used in the key generation, and the discrete Gaussian distribution $\chi_{err}$ is used to sample small noise polynomials. Specific details can be found in \cite{Bos} and \cite{Braserski}. The scheme is based on the following algorithms.
\begin{itemize}
\item \textbf{KeyGen}($n,q,t,\chi_{key},\chi_{err}$): Operating on the input degree $n$ and moduli $q$ and $t$, this algorithm generates the public and private keys $(pk,sk)=(h,f)$, where $f=[tf'+1]_{q}$ and $h=[tgf^{-1}]_q$. Here, the key generation algorithm samples small polynomials from the key distribution $f',g\rightarrow\chi_{key}$ such that $f$ is invertible modulo $q$ and $[.]_q$ denotes coefficients of polynomials in $R$ reduced by modulo $q$.

\item \textbf{Encrypt}($h,m$): Given a message $m\in R$, the Encrypt algorithm samples small error polynomials $s,e\rightarrow \chi_{err}$ and outputs, $c=[\lfloor q/t \rfloor[m]_t+e+hs]_q\in R$, where $\lfloor.\rfloor$ denotes the floor function.

\item \textbf{Decrypt}($f,c$): Given a cipher-text $c$, this algorithm outputs, $m=\left[\left\lfloor\frac{t}{q}[fc]_q\right\rceil\right]_t\in R$.

\item \textbf{Add}$(c_1,c_2)$: Given two cipher-texts $c_1$ and $c_2$, this algorithm outputs $c_{add}(c_1,c_2)=[c_1+c_2]_q$.

\item \label{Mult_Brase}\textbf{Mult}$(c_1,c_2)$: Multiplication of cipher-texts is performed in two steps. First, compute $\widetilde{c}_{mult}=\left[\left\lfloor\frac{t}{q}c_1c_2\right\rceil\right]_q$. However, this result cannot be decrypted to the original plain-text using the decryption key $f$. Therefore, a process known as key switching is done to transform the cipher-text so that it can be decrypted with the original secret key. For more details, we refer to \cite{Bos}.
\end{itemize}

This encryption scheme is homomorphic with respect to addition and multiplication of plain-texts modulo $t$. The main advantage in using Braserski's encryption scheme is that it can be used to encrypt blocks of plain-text instead of dealing with single bits, as in the DGHV scheme\cite{Dijk}. For example, consider the block of plain-text bits, $10100$. The integer representation of this block is the value $20$. We can represent this integer using the polynomial $X^2+X^4=\sum_{i=0}^{4}2^{i}z_i$, where $z_i$ is the $i$\textsuperscript{th} bit of $10100$.  In general, if $z$ is an integer and its binary representation is, $z=(\pm 1)\sum_{i=0}^{l}2^{i}z_i$, where $z_{i}\in\{0,1\}$ and $l=\lceil\log_{2}{|z|}\rceil$, then we can encode the number $z$ as $\sum_{i=0}^{l}z_{i}X^{i}\in R$. 

\subsection{Converting the plain-text space into a Field}
As we shall see, in our \textit{Homomorphic Query Processing} method, we invert certain plain-text elements and thus the plain-text space should be a field. Therefore, we now discuss how to convert the plain-text ring in Braserski's method to a field. Note that the plain-text space in Braserski's method is defined on the polynomial ring, $R_t=\mathbb{Z}_t[x]/\left<f(x)\right>$. We shall select $t=p$, where $p$ is a prime number. Then $R_p$ is a field if and only if $f$ is irreducible over $\mathbb{Z}_p$. Recall that $f$ is the $n$\textsuperscript{th} cyclomatic polynomial defined as follows:
\[f(x)=\prod_{\mathclap{\substack{1\leq k\leq n\\\gcd(k,n)=1}}}\left(x-e^{2i\pi\frac{k}{n}}\right)\]
Let $f(x)=(x-\alpha_1)(x-\alpha_2)\ldots(x-\alpha_n)$ be a polynomial defined on $\mathbb{Q}[x]$. The discriminant of $f$, denoted by $\Delta(f)$, is defined\cite{Harrison} as,
\[\Delta(f)=\prod_{i<j}(\alpha_i-\alpha_j)^2\]   
It has been proved in \cite{Harrison} that the $n$\textsuperscript{th} cyclotomic polynomial reduces modulo all primes if and only if the discriminant of the $n$\textsuperscript{th} cyclotomic polynomial is a square in $\mathbb{Z}$. Hence, by choosing a cyclotomic polynomial whose discriminant is not a square we can find a prime $p$ such that $f$ is irreducible over $\mathbb{Z}_p$. Furthermore, it is shown in \cite{Harrison} that whenever the discriminant of a cyclotomic polynomial $f$ is not a square in $\mathbb{Z}$, there exist infinitely many primes such that $f$ is irreducible over $\mathbb{Z}_p$. Thus, we can choose a cyclotomic polynomial with non-square discriminant and check for irreducibility using a standard polynomial irreducibility test such as Rabin's test, until we obtain a prime for which the cyclotomic polynomial is irreducible. For example, even if we consider a large cyclotomic polynomial with non-square discriminant like the 107\textsuperscript{th} cyclotomic (which has degree 106), and consider the primes less than 100, it can be seen that it is irreducible over many primes: $\mathbb{Z}_2,\mathbb{Z}_5,\mathbb{Z}_7,\mathbb{Z}_{17},\mathbb{Z}_{31},\mathbb{Z}_{43},\mathbb{Z}_{59},\mathbb{Z}_{67},\mathbb{Z}_{71},\mathbb{Z}_{73} \mbox{ and }\mathbb{Z}_{97}$. 

We now propose our Homomorphic Query Processing scheme, which is compatible with the Braserski's ring based fully homomorphic encryption scheme mentioned previously. 

\subsection{Homomorphic Query Processing}
\label{Gen_Gahi}
We begin by defining the value $F_i$ for the $i$\textsuperscript{th} record (denoted by $R_i$) in the database. We write $\mbox{Enc}(m)$ for the $\mbox{Enc}(m,pk)$, where $pk$ is the public key of the user. Then the user sends $\mbox{Enc}(m)$ to the server to search for the records that match $m$. Now, the server computes the following:
\begin{equation}\label{eq:Gen_Gahi}
F_i=\left(\prod \mbox{Enc}(m-R_k) \right)\left(\prod \mbox{Enc}\left(R_i-R_k\right)^{-1}\right),
\end{equation}
where each of the products above  is over all the records $R_k$ such that $R_k\neq R_i$. Since we are dealing with a fully homomorphic encryption scheme, we can compute $\mbox{Enc}(m-R_k)$ values by computing $\mbox{Enc}(m)-\mbox{Enc}(R_k)$. Also, since all the $R_i$ values are known to the server, the term $\prod_{R_k\neq R_i}\mbox{Enc}\left(R_i-R_k\right)^{-1}$ can be reduced to a simpler form using the homomorphic property of the encryption scheme in order to perform a single encryption. That idea of defining the $F_i$'s in this way  was inspired  by Lagrange Interpolating Polynomials. Indeed, we have 
\begin{align*}
F_i&=\mbox{Enc} \left( \prod_{R_k\neq R_i}(m-R_k)\right)\mbox{Enc}\left(\prod_{R_k\neq R_i}(R_i-R_k)\right)^{-1}\\
&=\mbox{Enc}\left(\prod_{R_k\neq R_i} \frac{m-R_k}{R_i-R_k}\right)
\end{align*}

We remark  that whenever $m=R_i$ (query being equal to the record we are comparing), we have $F_i=\mbox{Enc}(1)$ and $F_i=\mbox{Enc}(0)$, otherwise. Note that here we are assuming that the query is contained somewhere in the database. If the query is not contained anywhere in the database, an encryption of something other than 1 or 0 will be the output. This special scenario is discussed later. 

Now, we define  the partial sums of the $F_i$ values as follows: 
\begin{equation}\label{eq:Gen_Gahi2}G_i=\sum_{j\leq i}F_{j}.\end{equation}

Using these partial sums, we can then calculate the sequence $(F'_{i,k})$ corresponding to each record as follows,
\begin{equation}\label{eq:Gen_Gahi3}
F'_{i,k}=
F_i \left(\prod_{j\neq k} G_i-\mbox{Enc}(j) \right)\left(\mbox{Enc} \prod_{j\neq k}(k-j)^{-1}\right),
\end{equation}
where $1\leq k\leq i$. It can be seen that $F'_{i,k}=\mbox{Enc}(1)$ if $F_{i}=\mbox{Enc}(1)$ and $G_i=\mbox{Enc}(k)$ are both satisfied. Hence, the sequences $(F'_{i,k})$ have the property that whenever $F_{i}=\mbox{Enc}(1)$ (i.e., the $i$\textsuperscript{th} record matches the query), we have an $\mbox{Enc}(1)$ at the $k$\textsuperscript{th} position of the sequence where $G_i=\mbox{Enc}(k)$. All other entries of the sequence are encryptions of zero. Therefore,
\begin{equation}(R')=\sum_{k}\mbox{Enc}(R_{k})(F'_{k}),\label{eq:Mod_Gahi}\end{equation}
where $R_{k}$ is the $k$-th record in $D$ will give us a sequence containing only the encrypted records that match our search query. Note that the definition of $(R')$ relies on adding vectors of different lengths. This is done in the natural way, whereby all the vectors are made the same length by padding with zeros prior to addition.

To further illustrate our scheme, let us consider an example where the database contains five records, each with $4$ bits of data. Also, let our encryption scheme encrypt $2$ bits at a time. Then, if the search query is $(\mbox{Enc}(2),\,\mbox{Enc}(3))$, the corresponding $F_i$ and $G_i$ values are given in Table \ref{tab:Gen_Gahi}.

\begin{table}[ht]
\caption{Sample database and corresponding $F_i$ and $G_i$ values \label{tab:Gen_Gahi}}
\centering
\begin{tabular}{|c||c||c|}
\hline
Database Records & $F_i$& $G_i$\\
\hline
$(0,0,1,0)$ & $\mbox{Enc} (0)$&\mbox{Enc} (0)\\
$(1,0,1,1)$ & $\mbox{Enc} (1)$&\mbox{Enc} (1)\\
$(1,0,0,1)$ & $\mbox{Enc} (0)$&\mbox{Enc} (1)\\
$(1,0,1,1)$ & $\mbox{Enc} (1)$&\mbox{Enc} (2)\\
$(1,1,0,0)$ & $\mbox{Enc} (0)$&\mbox{Enc} (2)\\
\hline
\end{tabular}
\end{table}

The resulting sequences $(F'_i)$ would be similar as in Gahi's scheme,
\begin{align*}
(F'_{1})&=(\mbox{Enc}(0))\\
(F'_{2})&=(\mbox{Enc}(1),\mbox{Enc}(0))\\
(F'_{3})&=(\mbox{Enc}(0),\mbox{Enc}(0),\mbox{Enc}(0))\\
(F'_{4})&=(\mbox{Enc}(0),\mbox{Enc}(1),\mbox{Enc}(0),\mbox{Enc}(0))\\
(F'_{5})&=(\mbox{Enc}(0),\mbox{Enc}(0),\mbox{Enc}(0),\mbox{Enc}(0),\mbox{Enc}(0)).
\end{align*}
Therefore, the sequence $(R')$ would be,
\[(R')=(\mbox{Enc}(R_{2}),\mbox{Enc}(R_{3}),\mbox{Enc}(0),\mbox{Enc}(0), \mbox{Enc}(0))\]
At this point, the sequence $(R')$ will contain all the records that match our query but with trailing encryptions of zeros which we do not need. Hence, a second sum is calculated at the server side to determine the number of terms that are useful in the sequence:
\[n=\sum_{r} F_r.\]
Then $n$ will be returned to the user and decrypted to obtain the number of records that match the search query. Hence, the sequence $(R')$ can be truncated at the appropriate point and returned to the user for decryption. 

It should be noted that the server will know the number of records that match the user's query. We believe that this information is not sufficient for the server to gain any additional information about the search query. Alternatively, we could return the whole sequence without truncation, keeping the number of matching records private from the server. However, the communication overhead will be increased significantly in this case, since the length of the sequence will be equal to the number of records in the database.  

As promised previously, we now look at the special case where the record that is searched for is not contained anywhere in the database. In this case the value $F_i$ will be something other than an encryption of 1 or 0. These garbage encrypted values will carry themselves into the rest of the protocol, resulting in Equation \eqref{eq:Mod_Gahi} with a nonsensical sequence. Hence, if the user receives a nonsensical sequence as the final result, it implies that the record that was searched is not contained in the database. As an alternative approach, we can compute $\prod_i \left(\mbox{Enc}(m)-\mbox{Enc}(R_i)\right)$ prior to computing the $F_i$ in Equation \eqref{eq:Gen_Gahi} and send it to the user to decrypt. If the result is zero then $m$ is contained in the database, and if it is non-zero, $m$ is not contained in the database and therefore the user can send a message to the server to abort the search. 

\section{Comparison of Our Scheme vs. Gahi's Scheme}
Our scheme has the main advantage of having the potential to be used with more recent fully homomorphic encryption schemes rather than being restricted to the DGHV scheme. This gives the flexibility to use our method with block based encryption schemes such as Braserski's\cite{Braserski}, which reduces the number of encryption steps. For example, referring back to Equation \eqref{GahiEquation}, we can see that the $I_t$ values are calculated by comparing the query with each record bit-wise. If there are $m$ records in the database and each of them are encrypted using $n$ bits, the number of operations that are required to calculate all the $I_t$ values will be $\mathcal{O}(nm)$. In our Homomorphic Query Processing method, Equation \eqref{eq:Gen_Gahi} acts as the analogue of Equation \eqref{GahiEquation}. However, the encryptions are done block-wise in our scheme, and hence the number of operations it would take to calculate the $F_i$ value in Equation \eqref{eq:Gen_Gahi} will be $\mathcal{O}(m)$. For Equation \eqref{PartialSums} in Gahi's method, the number of operations that should be performed to calculate all the partial sums will be $\mathcal{O}(nm^2)$, since there are $\mathcal{O}(m^2)$ multiplications and each multiplication should be done bit-wise; whereas the calculation of partial sums in our scheme (Equation \eqref{eq:Gen_Gahi2}), the number of operations is reduced to $\mathcal{O}(m^2)$. Similarly, equations \ref{GahiEquation2} and \ref{GahiEquation3} in Gahi's method use $\mathcal{O}(nm)$ and $\mathcal{O}(nm^2)$ number of operations, respectively, but their counterparts in our scheme, (equations \ref{eq:Gen_Gahi3} and \ref{eq:Mod_Gahi}) have $\mathcal{O}(m)$ and $\mathcal{O}(m^2)$ operations, respectively. Thus, it can be seen that in each step of our scheme, the number of operations performed is reduced by a factor of $n$ compared to Gahi's method.

\end{document}